\documentclass[twocolumn,aps,showpacs,superscriptaddress,floatfix,
               preprintnumbers,amsmath,amssymb,nofootinbib]{revtex4}

\usepackage{graphicx}
\usepackage{dcolumn}
\usepackage{bm}

\begin{document}

\title{Impact of Low-Energy Constraints on Lorentz Violation}

\author{John Ellis}
\affiliation{Theory Division, CERN, CH 1211 Geneva 23, Switzerland}
\author{E. Gravanis and N.E.~Mavromatos}
\affiliation{Department of 
Physics (Theoretical Physics), King's College
London, Strand, London WC2R 2LS, United Kingdom}

\author{D.V.~Nanopoulos}\affiliation{Department of Physics, Texas A \& M University, 
College Station, TX~77843, USA; \\
Astroparticle Physics Group, Houston
Advanced Research Center (HARC), 
Mitchell Campus,
Woodlands, TX~77381, USA; \\
Academy of Athens, 
Division of Natural Sciences, 28~Panepistimiou Avenue, Athens 10679, Greece}

\begin{abstract} 

We extend previous analyses of the violation of Lorentz invariance induced
in a non-critical string model of quantum space-time foam, discussing the
propagation of low-energy particles through a distribution of
non-relativistic D-particles. We argue that nuclear and atomic 
physics experiments do
not constitute sensitive probes of this approach to quantum gravity
due to a difference in the dispersion relations for massive probes as 
compared to those for massless ones, predicted by the model.

\end{abstract} 

\maketitle

There has recently been considerable interest in possible violations of
Lorentz invariance, motivated theoretically by certain models of quantum
gravity~\cite{EMNfoam} and experimentally by high-energy cosmic-ray
data~\cite{GMCG}. Various tests of Lorentz invariance have been proposed,
including measurements of photons and other relativistic particles emitted
by astrophysical sources~\cite{Nature}, and recently low-energy tests in
atomic, nuclear and particle physics~\cite{SUV}. It is important to
compare the sensitivities of such non-relativistic tests with those of
relativistic probes, to see, for example, whether low-energy constraints
might rule out the observation of significant dispersion in the arrival
times of photons from gamma-ray bursters (GRBs). A first discussion of
low-energy constraints in this context has recently been given
in~\cite{SUV}, where it was argued that nuclear physics constrains certain
Lorentz-violating parameters to be much smaller than might have been
suggested by simple dimensional analysis and the Planck length $\ell_P
\sim 10^{-33}$~cm.

Three of the present authors (JE, NEM, DVN) have proposed previously a
dynamical model of Lorentz violation within the framework of non-critical
string theory~\cite{EMNfoam}, in which the Liouville string mode is
interpreted as the target time variable. In this approach, the recoil of a
point-like D-brane defect (called a D-particle from now on)  in the
quantum space-time foam, when struck by a passing closed-string matter
particle state, has a non-trivial back-reaction on the surrounding
space-time, modifying the effective metric felt by the particle and hence
its propagation. This effect results in a mean-field four-dimensional
metric of non-diagonal form, providing a mechanism for Lorentz violation
in non-critical string:

\begin{equation}
\label{foammetric}
G_{0i} = v_i~, ~\qquad G_{ij}=\delta_{ij}, ~\qquad G_{00}=-1,
\end{equation}
where $v_i \sim P_i / M$, with $P_i$ the particle momentum and $M (\sim
M_P ?)$ the effective mass of a D-brane defect in the quantum space-time
foam. This 
results 
in a
{\it reduction} in the velocity of the propagating particle:
\begin{equation}
u \sim c (1 - P/M).
\label{modifyc}
\end{equation}
In this paper we give a more elaborate derivation of this 
relation 
and extend 
it to massive non-relativistic particles,
within the framework of our Liouville formalism. 
To this end, we first discuss the main relevant
aspects of our approach based on world-sheet field theory. 

We assume that the dynamics of propagating particles is described by some
critical string theory as long as their interactions with quantum foam
fluctuations in the latter may be neglected, and we model space-time foam
as a gas of D-particles. Interactions with these cause the effective
theory describing the propagating states to become a non-critical string
theory, which we treat using the formalism of a renormalizable $\sigma$
model on the world-sheet. The recoil velocities of the struck D-particles
constituting the foam are viewed~\cite{EMNfoam} as couplings in this
$\sigma$ model that are not exactly marginal. The non-criticality of this
effective string model is compensated by non-trivial dynamics of the
string Liouville mode, which we interpret as the target time variable.

It is important to distinguish this approach from {\it ad hoc}
phenomenological modifications of Lorentz-invariant dispersion relations
in {\it flat } space times. Our approach entails a transition between
conformal field theories (CFTs) on the world sheet: one starts from a
system of a D-particle interacting with closed strings, which defines the
asymptotic past conformal field theory (CFT1). Long after the scattering
at $t=0$ say, the D-particle is moving with constant velocity, and the
system is described by a different future CFT (CFT2).  The transition
between these two CFTs is described in a mathematically consistent way by
Liouville string. The introduction of the latter requires an extra
space-time (Liouville) field, of time-like signature~\cite{EMNfoam}. This
intermediate $(D+1)$-dimensional target space is {\it curved}, as a result
of Liouville dressing and the presence of the boundary operator describing
recoil of the D-particle. At asymptotically long times after the
scattering event, the space-time becomes flat, but it differs via the
off-diagonal constant metric terms in (\ref{foammetric}), which arise from
the identification of the world-sheet zero mode of the Liouville field
with the target time.  For this reason, it is not trivial to describe the
effective low-energy dynamics simply in terms of an equilibrium
flat-space-time field theory with naive violations of Lorentz
symmetry~\cite{SUV}. 

The metric (\ref{foammetric}) arises from the recoil of an initially
stationary defect after it is struck by a light closed-string state, in
the the semi-classical approximation. In a world-sheet framework, this
implies a restriction to world-sheet surfaces with trivial topologies. For
open-string excitations on the D-brane, which interest us here, this means
a world-sheet with disc topology.  The relevant recoil deformations of the
$\sigma$-model action take the form~\cite{kmw,szabo}:

\begin{equation}
{\cal V}_{\rm rec} = \int _{\partial \Sigma} \theta_\epsilon (X^0)
\left(\epsilon^2 y^i + \epsilon v^i X^0 \right)\partial_{\rm n} X^i 
\label{recoildef}
\end{equation}
where $\partial \Sigma$ is the boundary of the world-sheet
disc, $\partial_{\rm n}$ denotes the derivative normal to the world-sheet,
$y^i$ denotes the initial position of the D-particle, and $v^i$ its 
recoil velocity after the scattering by a closed-string state.
The regulating parameter $\epsilon$, which serves to regulate the 
Heaviside function $\theta(X^0)$, is related to the world-sheet 
renormalization-group scale by
$\epsilon^{-2} \sim {\rm ln}(L/a)^2$, in order to close the logarithmic 
conformal 
algebra characterizing the recoil~\cite{kmw,szabo}. 
The couplings $y^i, v^j$ written in (\ref{recoildef}) 
are exactly marginal, i.e., independent of  the scale $\epsilon$~\cite{szabo}.

The above considerations were in a frame in which the 
D-particle was {\it at rest}. We now extend the discussion to motion 
through a gas of moving D-particles, as is likely to be the case for
a laboratory on Earth, e.g., if the D-particle foam is comoving 
with the Cosmic Microwave Background (CMB) frame. 
Assuming this to be moving with three-velocity ${\vec w}$ 
relative to the observer,
the recoil deformation takes the following form
to leading order in $\epsilon \to 0^+$:
\begin{eqnarray}
&&{\cal V'}_{\rm rec} 
= \int _{\partial \Sigma} 
\theta_\epsilon (-X^0_w)\gamma_{\epsilon w} 
\epsilon w^i X_w^0\partial_{\rm n} X^i_w
+  \nonumber \\
&& \int _{\partial \Sigma} 
\theta_\epsilon(X^0_w)
\gamma_{\epsilon w} \epsilon v^i(\epsilon, w))X_w^0 
\partial_{\rm n} X^i_w
\label{changerec}
\end{eqnarray}
where the suffix $w$ denotes quantities in the boosted frame. 
The main novelty in the $w \ne 0$ case 
is that now there are two $\sigma$-model operators in (\ref{changerec}). 
The recoil velocity $v^i(\epsilon, w)$ 
depends in general on $w$, and is determined by momentum
conservation during the scattering process, as discussed in~\cite{szabo}:
$\epsilon v_\parallel (\epsilon, w) = 
(1 + \epsilon ^2 v \cdot w)^{-1}\left(\epsilon v + \epsilon w \right)$, and 
similarly for the $v_\perp$ component. 
Finally, $\gamma_{\epsilon w} \equiv 1/\sqrt{1- \epsilon^2 w^2}$. 
The above formulae have been derived by applying the standard Lorentz 
composition of velocities to the bare, i.e., non-marginal, couplings
$\epsilon v,\epsilon w$. This is justified because Lorentz 
transformations are 
classical changes of coordinates, and as such should be applicable to
bare quantities appearing on the $\sigma$ model. After 
renormalization at the $\sigma$-model level~\cite{szabo} 
the marginal couplings
$u^i,w^i$ obey, to leading order in $\epsilon$, a {\it Galilean composition}.
In this way the higher-order terms, 
particularly those of order $w^2$, which were 
crucial in the analysis of~\cite{SUV}, are suppressed by 
factors of order $\epsilon^2$, which, as we shall discuss below, are relaxation
terms, scaling with time as $1/t$. 
In this way, all dangerous Lorentz-violating terms, which would be
severely constrained by low-energy  data as discused in~\cite{SUV}, relax 
in this way, and therefore are {\it suppressed} in our Liouville model. 
As we show below, this implies that, at present, the most 
sensitive probe of Liouville-gravity-induced quantum effects 
is that associated with studies of astrophysical sources such as gamma-ray 
bursters~\cite{Nature,Sakharov}. 

The deformations (\ref{recoildef}) and (\ref{changerec}) are relevant 
world-sheet deformations in a two-dimensional renormalization-group 
sense, with anomalous scaling dimensions $-\epsilon^2 /2$~\cite{kmw}. 
Their presence drives the stringy $\sigma$ model non-critical,
and requires dressing with the Liouville mode $\rho$. 
The procedure has been described in detail for the $w=0$ case
in~\cite{EMNfoam}, so we are brief in what follows. 
In Liouville strings there are two screening operators 
$e^{\alpha_\pm\rho}$, where the 
$\alpha_\pm$ are the Liouville-string anomalous dimensions
given by:
\begin{equation} 
\alpha_\pm = -Q/2 \pm \sqrt{\frac{Q^2}{2} + \frac{\epsilon^2}{2}}
\end{equation} 
and the central-charge deficit $Q^2$ was computed in~\cite{EMNfoam} and 
found to be of order $\epsilon^4$. Hence $\alpha_\pm \sim \pm \epsilon$. 

The $\alpha_-$ screening operator is sometimes neglected because it
corresponds to states that do not exist in Liouville theory. However, this
is not the case in string theory, where one should keep both screenings as
above. This is essential for recovering the correct limit of vanishing
recoil: $v^i \to 0$ in the case of infinite D-particle mass.
The Liouville-dressed boosted deformation reads (to leading order 
in $\epsilon \to 0^+$): 
\begin{eqnarray}
&&{\cal V'}^L_{\rm rec} =
\int _{\partial \Sigma}e^{\alpha_-\rho}  
\theta_\epsilon (-X^0_w)\epsilon w^i X_w^0 \partial_{\rm n} X^i_w
+  \nonumber \\
&&\int _{\partial \Sigma} e^{\alpha_+\rho}  
\theta_\epsilon(X^0_w)
\epsilon (w^i+ v^i)X_w^0\partial_{\rm n} X^i_w
\label{liouvdressed}
\end{eqnarray}
where $v^i$ is the recoil velocity in the frame where the D-particle 
in initially at rest.  
Using Stokes' theorem, and ignoring terms that vanish
using the world-sheet equations of motion, one arrives easily at
the following bulk world-sheet operator:
\begin{eqnarray} 
{\cal V'}^L_{\rm rec} =
\int _{\Sigma}e^{\epsilon \rho}  
\theta_\epsilon (X^0_w)\epsilon^2  
v^i X_w^0\partial_{a} X^i_w\partial^a \rho 
+ \dots,~a=1,2,  
\label{liouvdressedbulk}
\end{eqnarray}
where the $\dots $ denote terms subleading as $\epsilon \to 0^+$, as we 
explain below. Notice the {\it cancellation} of the terms proportional to 
$w$, due to the opposite screening dressings. Recalling that the 
regularised Heaviside
operator $\theta_\epsilon (X) = \theta_0 (X) e^{-\epsilon X}$, where
$\theta_0 (X)$ is the standard Heaviside function, we observe that one can
identify the boosted time coordinate $X_w^0$ with the Liouville mode
$\rho $: $\rho = X_w^0$~\cite{EMNfoam}. At long times after the 
scattering, the Liouville-dressed theory leads to 
target-space metric deformations of the following form to order 
$\epsilon^2$: 
\begin{equation}\label{metricdressed} 
G_{\rho i} = 
\epsilon^2  v^i \rho ~+~{\rm relaxation~terms} 
\end{equation}
As explained in detail in~\cite{kmw,EMNfoam}, at the long times 
after the scattering event when the $\sigma$-model formalism
is valid, one has $\epsilon^2 \rho \sim 1$: $\epsilon$ and
the world-sheet zero mode of $\rho$  are not independent
variables, as $\epsilon$ is linked
with the world-sheet renormalization-group scale.  
Thus we recover the metric (\ref{foammetric}). 
Notice that 
the 
above-described 
$\sigma$-model 
formalism, which was invented for a first order analysis in the small recoil 
velocity~\cite{kmw}, keeps only linear velocity $w$ terms in the  Lorentz 
transformation, which is mathematically self consistent. 
The reader should keep in mind that 
the recoil velocity $v$ is the momentum transfer to the D-particle, and thus
its first correction under a Lorentz transformation is of order $w^2$.
In particular, this implies that this 
specific $\sigma$-model framework is not tailored to  
give a definite answer to tests sensitive to second order in $w$~\cite{SUV}.
Nevertheless, as we shall argue below, there are ways of tackling this problem
upon making a few physically reasonable assumptions.

To this end, we first notice that in our $\sigma$-model framework,
the metric is actually a field operator, cf. (\ref{liouvdressedbulk}),
\begin{eqnarray} 
&~& G_{\rho i}(z,{\bar z})= 
\epsilon v^i 
:e^{\epsilon \rho(z,{\bar z})}\theta_\epsilon (X^0(z,{\bar z}))X^0(z,{\bar z}):
\equiv \nonumber \\
&~&\epsilon v^i:e^{\epsilon \rho} {\cal D}:~, 
\label{gmetr} 
\end{eqnarray} 
where :...: denotes normal ordering, and ${\cal D}$ is the recoil velocity  
operator of \cite{kmw}, which obeys a logarithmic conformal algebra. 
In our case, $G^{\mu\nu}=\eta^{\mu\nu} 
+ h^{\mu\nu}$, and $G_{\rho i} = h_{0i}$ (eventually $\rho$ is identified
with the temporal coordinate). 
One should average 
the above dispersion relation  
with respect to the Liouville $\sigma$-model action
\begin{equation} 
\langle p_\mu p_\nu G^{\mu\nu}\rangle= p_\mu p_\nu \eta^{\mu\nu} - 2E p_i
\langle h_{0 i} \rangle = -m^2~,
\label{average} 
\end{equation} 
Due to normal ordering,
one obtains immediately $\langle h_{0i} \rangle =0$.

However, this is not the end of the story, as one should consider
$\sigma$-model two-point correlations appearing in the average of 
the square of the dispersion relation. In general, this  
approach is equivalent to the standard approach of considering 
fluctuations about a mean value in stochastic frameworks
of quantum gravity~\cite{stoch}. 
Such issues will be discussed further in~\cite{gravmav}. 

Squaring the dispersion relation and taking the above average we have: 
\begin{eqnarray}
&~& m^4 = p_\mu p_\nu p_\alpha p_\beta \langle G^{\mu\nu} G^{\alpha\beta} \rangle 
= \nonumber \\
&~& (E^2-p^2)^2 + E^2 p_i p_j \langle h_{0i} h_{0j}\rangle~.
\label{square}
\end{eqnarray} 

In our case we can use Liouville $\sigma$-model methods to compute the 
above correlator, taking into account the fact that in our approach 
we identify the world-sheet zero mode of the Liouville field $\rho_0$ 
with that of the field $X^0$. Splitting the Liouville 
path integration~\cite{ddk} into zero-mode ($\rho_0$) and non-zero-mode 
parts (${\tilde \rho_0}$), we obtain: 
\begin{eqnarray} 
&~& \langle h_{0i} h_{0j}\rangle 
= \epsilon^4 v_i v_j \int d\rho_0~{\rm exp}[-\epsilon^2 \chi \rho_0 + \dots ]
\nonumber \\
&~& \int D{\tilde \rho}~{\rm exp}
\left(-\int_\Sigma (\partial {\tilde \rho})^2 + \epsilon^2 \int_\Sigma R^{(2)}\rho + \dots \right) \nonumber \\
&~& \langle\langle e^{\epsilon \rho} {\cal D}(z,{\bar z}) 
 e^{\epsilon \rho} {\cal D} (z,{\bar z})\rangle\rangle \sim 
\epsilon^4 \frac{v_iv_j}{\epsilon^2} \frac{1}{\epsilon^2} \times ({\rm finite}) 
\label{correlator} 
\end{eqnarray} 
where $\Sigma$ is the world sheet of curvature $R^{(2)}$ and 
Euler characteristic 
$\chi$, and 
$\langle\langle \dots \rangle\rangle $ denote the $\sigma$-model 
path integral over the $X^\mu$ fields. 
Above we took into account the fact that in our case the 
square root of the central-charge deficit 
is of order $\epsilon^2$. With these in mind, as well as the fact that
the zero mode of the Liouville field is related to the logarithm of the 
world-sheet area, we could transform the Liouville zero-mode integral
to an area integral by inserting the fixed area constraint~\cite{ddk},
which yields eventually an $1/\epsilon^2$ divergence. 
A further $1/\epsilon^2$ divergence is obtained from the logarithmic 
algebra of the ${\cal D}$ operator~\cite{kmw,szabo}. The non-zero mode
in the Liouville integral yields finite results, as can easily be seen. 
In addition to the tree-level $\sigma$-model computation, one
should consider contributions of higher genera~\cite{szabo}, 
which renormalize the 
leading result but do not change it qualitatively~\cite{gravmav}. 
The renormalization procedure involved in the computation of the 
two-point correlator when string loops are included 
above may change its (apparently) positive sign~\cite{gravmav},
due to subtractions. In what follows we shall assume that the sign is negative. 

The dispersion relation obtained in this way reads:
\begin{equation} 
m^4=(E^2-p^2)^2 -\xi^2 E^2 (p_i v^i)^2 
\label{disp}
\end{equation} 
where $\xi$ is a number. 
In our case $ v^i \sim g_s p^i/M_s$ and hence $M=M_s/g_s$, 
where $g_s$ is the 
string coupling, assumed to be weak, and $M_s$ is the string
scale, which may in general be different from the Planck scale 
$M_P$.

In the case of a massless particle such as a photon (or at high 
energies: $m/E \to 0$) the above dispersion relation yields: 
\begin{equation} 
E^2 = p^2 \pm \xi g_s \frac{p^3}{M_s} + \dots 
\label{massless}
\end{equation} 
Subluminal dispersion relations are expected in the case of string 
theory, motivating the negative sign. This stems from
the specific form of the low-energy target-space dynamics describing the
recoil, which takes the form of a Born-Infeld action~\cite{szabo,EMNfoam}.
This yields a refractive index that is linear in $E$, minimally 
suppressed by one power of the Planck scale~\cite{Nature,EMNfoam}.
The above procedure of considering two-point correlators 
yields stochastic fluctuations in the arrival times of 
photons~\cite{EMNstoch}, independent from the modification of the
dispersion relation. If the two-point correlator  
turned out to be positive, only transverse fluctuations would exist, 
implying that there would be 
no modifications in the photon's dispersion relation.
However, the arrival-time fluctuations, expressing 
light-cone fluctuations~\cite{stoch,EMNstoch}, which are proportional to 
the square root of this correlator,
would still exist.

We now consider low-energy massive particles, as considered in~\cite{SUV}. 
Since $p \ll m \ll M_P$ in this case,
the dispersion relation takes the form: 
\begin{equation} 
E^2 = m^2 + p^2 + \frac{\xi^2}{2}\frac{g_s^2}{M_s^2}p^4 + \dots 
\label{massive} 
\end{equation}
Notice the {\it qualitative difference in the scaling with $M_s^2$ in this 
case}, which reduces drastically the sensitivity of the model to 
tests using massive low-energy
particles. Indeed, the last term, when applied to non-relativistic
fermions, as in~\cite{SUV}, would yield a quadrupole-moment term of order
$\xi^2 g_s^2 (m^2/M_s^2) ~{\vec w} \cdot Q \cdot {\vec w} $, which is
suppressed  by a factor 
$g_s^2 m/M_s $ as compared to the result of~\cite{SUV}, rendering it
unobservable in practice.

Before closing we would like to make an important remark.
So far we have considered recoil in a single string-D-particle scattering.
One may argue, however, on the possibility of having, instead 
of a single string, a 
{\it beam} of incident strings with some distribution 
$v(y)$ of 
velocities, with $y$ a direction transverse to that of propagation. 
In this case there are non trivial space-time 
curvature effects induced by the metric 
(\ref{metricdressed}) with $v_i $ replaced by 
the distribution $v_i(y)$, representing the mean field. 
The presence of such effects 
guarantees the impossibility of performing a 
coordinate transformation to remove the mean field result.

Thus, the induced metric now reads:
\begin{equation} 
G_{\mu\nu} = G_{\mu\nu}^0 + \frac{1}{3}R_{\mu\rho\nu\sigma}\delta x^\rho \delta x^\sigma
+ \dots~,
\label{curvmetric}
\end{equation} 
where the $\dots$ denote higher derivative terms, 
and $G_{\mu\nu}^0=G_{\mu\nu}(v(y=0))$ is the metric at, say, the center of 
the beam of particles. The induced curvature is of 
order: 
$R_{\mu\nu\rho\sigma} \sim \frac{v_i^2}{\ell^2}$, where $\ell$ is defined as the characteristic 
scale for the change of $v$: $\partial v/\partial y \sim v/\ell$. 

Therefore the mean field result of the dispersion relation 
is the standard one (in the frame where the initial D-particle was at rest):
\begin{equation} 
m^2 = E^2 -p^2 + \xi Ep_iv^i + {\cal O}(p^2v^2) 
\end{equation}
In case the  
D-particle moves with velocity $w$ with respect to the observer, 
one obtains  
Lorentz-violating quadrupole terms 
of order 
$\frac{g_s\xi}{M_s}w \cdot Q \cdot w $.  
In that case, the analysis of \cite{SUV} would imply a bound 
$\xi g_s M_P/M_s \sim 10^{-5}~$ which is not a very strong bound 
for our stringy case, 
where we have three parameters $\xi,~M_s/M_P$ and $g_s$.   
Moreover, in the case of a 
beam of non relativistic particles, with average velocity 
$p_i/m \ll 1$, there is an extra suppression factor 
in the Lorentz-violating term 
of order 
${\cal O}(p/m)$ due to the probability of scattering with the D-particle. 
However, for the nucleon energies (of 40 MeV) considered in  
\cite{SUV} the constraint is reduced by at most one order of magnitude.

Nevertheless, the mean field may not
be appropriate for a proper $\sigma$-model analysis, 
as it deals with 
non conformal (non-Ricci flat) metric backgrounds, and moreover
the formalism is not developed to discuss properly 
more than single string-D-particle interactions. 
Hence 
we believe that such constraints can be avoided within a
mathematically self-consistent Liouville $\sigma$-model framework
where the mean field is absent, 
but there are fluctuations, 
as described above.

This completes our discussion of relevant properties of our D-particle 
model
for space-time foam. 
As shown here,
the nuclear-physics constraint of~\cite{SUV}, as well as other 
low-energy experiments, by no means exclude $M \sim 10^{19}$~GeV.
For comparison, the latest astrophysical limit is $M \sim 7 \cdot
10^{15}$~GeV~\cite{Sakharov}. We recall also that an important difference 
of our foam model from other approaches to modified dispersion relations,
such as loop gravity, is that we have {\it no} superluminal signals. 
Models with superluminal propagation are essentially excluded by the
absence of gravitational {\v C}erenkov radiation from ultra-relativistic
particles~\cite{nelson}. Our Liouville string model escapes~\cite{escape}
from this and other constraints that severely restrict generic
Lorentz-violating models of quantum gravity, such as the phenomenology of
neutrino oscillations~\cite{lisi}. These properties occur
for specifically stringy reasons. Moreover, our dispersion
relations derived in \cite{volkov} and here 
are distinct from those proposed for
fermions in the context of the loop-gravity approach~\cite{alfaro}, thus
avoiding also constraints from cosmic-ray decays.

In our previous work~\cite{EMNfoam}, we have assumed that space-time is
populated by ${\cal O}(1)$ stationary defect per Planck volume, in which
case the rate of collisions of a relativistic particle is also ${\cal
O}(1)$ in natural units and, as mentioned above, 
a slow-moving particle would encounter fewer
defects, by a factor $\propto |p_i/m|$, suppressing the Lorentz violation
effect by a similar extra factor.  
The above analysis assumed
also unrealistically that all the D-particles in the foam have the same
velocity $w_i \ll c$. 
In general, one would expect a gas of moving D-particles with a
distribution of velocities ${\cal P}(w_D)$ that are different from the CMB
velocity. The discussion of such an ensemble goes beyond the scope of this
paper, but, as long as the $w_D$ are non-relativistic, the above analysis
would go through with $w \to \langle w_D \rangle$, and $\langle w_D
\rangle$ could be identified with the Earth's motion $w$ relative to the
CMB frame.  More generally, one could consider the possibility that the
distribution ${\cal P}(w)$ might extend to relativistic D-particle
velocities. A detailed treatment of this case is not possible within our
current calculational framework, but we would expect it to lead to similar
conclusions.

\section*{Acknowledgements} 

This work is supported in part by the European Union through contract
HPRN-CT-2000-00152. The work of E.G. is supported by a King's College
London Research Studentship (KRS). The work of D.V.N. is supported by
D.O.E. grant DE-FG03-95-ER-40917. J.E. thanks the Yukawa Institute at
Kyoto University for hospitality while this work was being completed.

\end{document}